\documentclass[conference]{IEEEtran}
\IEEEoverridecommandlockouts
\usepackage{cite}
\usepackage{amsmath,amssymb,amsfonts}
\usepackage[pdftex, dvipdfmx]{graphicx}
\usepackage{textcomp}
\usepackage{url}
\usepackage{algorithmic}
\usepackage[ruled,vlined]{algorithm2e}
\newcommand{\mi}[1]{\mathit{#1}}

\renewcommand{\baselinestretch}{0.955}
\def\BibTeX{{\rm B\kern-.05em{\sc i\kern-.025em b}\kern-.08em
    T\kern-.1667em\lower.7ex\hbox{E}\kern-.125emX}}
\begin{document}

\title{Trail: A Blockchain Architecture for Light Nodes\\
}
\author{
    \IEEEauthorblockN{Ryunosuke Nagayama, Ryohei Banno$^{\dagger}$, and Kazuyuki Shudo}
    \IEEEauthorblockA{\textit{Tokyo Institute of Technology} \\
    Tokyo, Japan \\
    nagayama.r.ac@m.titech.ac.jp}
}

\maketitle

\begin{abstract}
In Bitcoin and Ethereum, nodes require large storage capacity to maintain all the blockchain data, such as transactions, UTXOs, and account states.
As of May 2020, the storage size of the Bitcoin blockchain has expanded to 270 GB, and it will continue to increase.
This storage requirement is a major hurdle to becoming a block proposer or validator.
Although many studies have attempted to reduce the storage size, in the proposed methods, a node cannot keep all blocks or cannot generate a block.
We propose an architecture called Trail that allows nodes to hold all blocks in a small storage and to generate and validate blocks and transactions.
Trail does not depend on a consensus algorithm or fork choice rule.
In this architecture, a client who issues transactions has the data to prove its own balances and can generate a transaction containing the proof of balances.
The nodes in Trail do not store transactions, UTXOs and account balances: they keep only blocks.
The blocksize is approximately 8 KB, which is 100 times smaller than that of Bitcoin.
Further, the block size is constant regardless of the number of accounts and the number of transactions.
Compared to traditional blockchains, clients who issue transactions must store additional data. 
However, we show that proper data archiving can keep the account device storage size small.
Trail allows more users to be block proposers and validators and improves the decentralization of the blockchain.
\end{abstract}

\begin{IEEEkeywords}
blockchain, Bitcoin, Ethereum, miner, storage
\end{IEEEkeywords}

\renewcommand{\thefootnote}{\fnsymbol{footnote}}
\footnote[0]{$\dagger$ Current affiliation as of June 2020: Kogakuin University}
\renewcommand{\thefootnote}{\arabic{footnote}}

\section{Introduction}
Blockchain is a distributed system with Byzantine fault tolerance.
Because blockchain can manage a distributed ledger without a centralized system and makes tampering with past data difficult, blockchain is used as the core technology for cryptocurrencies.
In major blockchains such as Bitcoin and Ethereum, nodes validate received transactions and generate a block from valid transactions.
The node then broadcasts the block, and the receiving node validates the block and the transactions contained in the block.
In the transaction validation process, nodes validate that the sender has the balance to be remitted in the transaction.

In Bitcoin\cite{b0,b8}, the balance is managed using unspent transaction output (UTXO), which
contains the address of the owner and the amount of coins.
Bitcoin transactions contain UTXOs as input and assign the total balance of the input to the new UTXOs.
Since a UTXO must be used only once as a transaction input, nodes check whether the UTXO was used in past transactions;
therefore, Bitcoin nodes must store all past transactions and UTXOs.

In Ethereum\cite{b2,b11}, account balances are stored in a Merkle Patricia trie.
Unlike Bitcoin, there is no need for past transactions to validate the double use of coins.
However, since  fraud occurs when using the same transaction multiple times, nodes check whether the transaction has been approved in the past blocks.
Therefore, Ethereum nodes need to maintain account status information, such as balances and number of past transactions.

The data size of the blockchain is enormous because it includes all transactions or balances of all accounts.
As of May 2020, the storage size has expanded to 270 GB, and it will continue to increase\cite{b1}.
The large storage requirements make it difficult to become a node.
As the number of nodes increases, the decentralization and security of the blockchain improve;
therefore, the storage size of the blockchain is one of the important issues.

We propose the Trail architecture in which
account balances are managed in the same way as UTXOs
using the transaction output (TXO).
TXO is stored in a data structure called a TXO tree.
The TXO tree is used to manage whether a TXO is used or unused and transactions contain Merkle proof of TXOs; thus, nodes do not require past transactions and TXOs for validation.
Further, nodes can generate a block from only the parent block and new transactions;
therefore, the storage size of nodes is small.

Trail architecture has the following advantages:
\begin{itemize}
    \item Nodes do not have to keep transactions, UTXOs and account balances: they keep only blocks.
    \item Users can prove their balance to the other party without relying on nodes.
    \item The block size is constant regardless of the number of accounts and the number of transactions.
    \item Trail does not depend on the consensus algorithm or the kind of fork choice rule.
\end{itemize}

\section{Related Work}
This section describes existing research that attempts to reduce the storage size.

Nakamoto\cite{b0} proposed to prune the Merkle tree.
In Bitcoin, transactions that have been buried under a sufficient number of blocks are difficult to overturn.
Therefore, nodes save storage by summarizing old transactions into a hash value of a parent node and discarding the transactions themselves.
L. Quan et al.\cite{b5} analysed the distribution of the period from the approval of a UTXO to its use, and based on the analysis, proposed a method of discarding transactions properly.
In these methods, the node needs to hold the transactions for a certain period of time, and the node decides when to discard them.
In Trail, nodes do not need to keep any transactions, and balance management and data discarding are decided by the owner of the balance.

Simplified payment verification (SPV)\cite{b0} and light client~\cite{b3} are methods being researched by the Bitcoin and Ethereum communities.
These methods allow clients to validate blocks using the Merkle proof of transactions.
However, these methods still require block proposers to store transactions and account balances, whereas in Trail, block proposers do not need to keep transactions and account balances.
Further, clients in Trail keep the Merkle proof of their balances, so they can validate blocks in the same way as these methods.

Stateless client\cite{b7} is a method in which validators need to keep only blocks containing the tree root.
Stateless clients can validate blocks in the current research but cannot generate blocks.
In Trail, nodes can validate and generate blocks without keeping data other than blocks by adding information required for verification and generation of a block to the transactions.

Omniledger\cite{b12} reduces storage and improves throughput by sharding.
Each shard is randomly assigned to validators periodically.
When the client issues transactions between different shards, the associated shards commit or abort the transaction, and the aborted transaction is rolled back on each shard.
Ethereum will also implement sharding\cite{b4}.
Validators are randomly assigned, and the shard periodically commits blocks to the beacon chain, the blockchain that manages all shards.
Pegged Sidechain\cite{b10} is a method used to create another blockchain called Sidechain, which issues currency that can be exchanged for Bitcoin Blockchain.
A transaction is issued in two blockchains, and the transaction is validated using SPV.
This method reduces the storage by dividing the ledger and improves the throughput, similarly to sharding.

Vault\cite{b6} is a method to reduce storage and rapidly synchronize blockchains.
Vault defines the period during which a transaction can be included in a block.
If a block outside this period approves the transaction, the transaction is considered invalid.
Therefore, Vault nodes do not need to store expired transactions.
In Vault, the balance of each account is stored in a Merkle tree, and each node is assigned and holds a part of the Merkle tree.
By adding a Merkle proof to a transaction, a node can validate and generate a block even if the node keeps only a part of the Merkle tree.
These techniques allow nodes to reduce the number of transactions and the size of the Merkle tree they store.
By contrast, in Trail, nodes do not keep transactions: they store only the root and one Merkle proof of the Merkle tree per block by using the TXO tree.
Vault requires nodes to keep a part or all of Merkle tree in order to insert new leaf nodes into the Merkle tree.

Since the insertion position is predetermined in Trail, nodes need to keep only one Merkle proof.

\section{Asset Ownership Management on Blockchain}
Two main methods of ownership management of assets on the blockchain exist: account based and UTXO based.
Ethereum is an account-based blockchain.
Account-based blockchains record the assets of all accounts in each block:
the account and its assets are recorded in the block as one node in the Merkle tree.
Validators check the past block's record to confirm if the account holds the balance sent in the transaction.

By contrast, Bitcoin is UTXO-based blockchain.
A UTXO records the owner's address and the amount of coins.
An account holds one or more UTXOs, and the total amount of UTXOs represents the total assets of the account.
\begin{figure}[t]
    \centering
    \includegraphics[width=4cm]{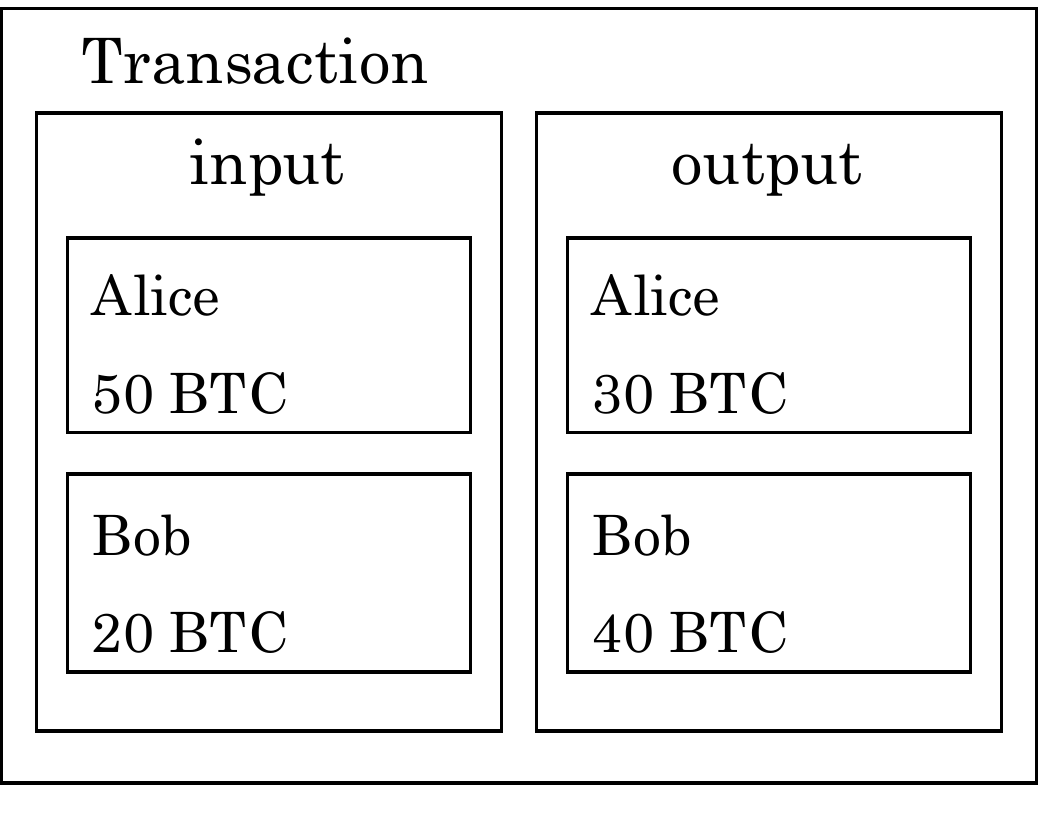}
    \caption{A transaction sending 20 BTC from Alice to Bob.}
    \label{fig:tx}
\end{figure}

As shown in Fig.\ref{fig:tx}, a transaction consumes one or more UTXOs as input and creates one or more new UTXOs as output.
This transaction consumes Alice's 50 BTC UTXO and Bob's 20 BTC UTXO and creates Alice's 30 BTC UTXO and Bob's 40 BTC UTXO.
If the transaction is approved, Alice's assets decrease from 50 BTC to 30 BTC and Bob's assets increase from 20 BTC to 40 BTC.
As a result, 20 BTC is sent from Alice to Bob.
The approved transaction is recorded in the block as one node in the Merkle tree.
A validator checks the past block to confirm if the UTXO of the transaction input has been used as the transaction input in the past;
it is illegal to use a UTXO as transaction input multiple times.

Trail is UTXO based;
however, it records whether a UTXO was used as input in the past in a different way than that in Bitcoin.
Bitcoin creates a Merkle tree with transactions as leaf nodes and records the tree in blocks, while Trail records Merkle trees with UTXOs as leaf nodes in blocks.

\section{Overview of Trail Architecture}

This section describes the design of Trail architecture.
In the following, we use the terms "Trail node" and "client".
A client simply issues a transaction.
A Trail node validates transactions and blocks and generates blocks.
A Trail node is usually also a client.
Moreover, the nodes in the tree structure are simply called "nodes".

Since clients include witnesses of their assets in transactions, Trail allows Trail nodes to validate transactions and generate blocks with minimal data storage.
A simulator implementing the Trail architecture has been released on Github\footnote{\url{https://github.com/nagayamaryu/trail_simulator}}.

\subsection{TXO Tree}
First, we describe the TXO tree, which is the core idea of Trail.
The TXO tree is a Merkle tree with leaf nodes that store the hash values of all TXOs approved in the past blocks on the blockchain in such a way that it can be determined whether a TXO has been used.
In Bitcoin, a new Merkle tree is created for each block, while Trail updates one TXO tree through the blockchain and records the snapshot of the TXO tree in the block:
no one needs to store the entire TXO tree.
A TXO and its Merkle proof are held only by the client who is owner of the TXO, and the Trail node holds only the root of the TXO tree, one hash value and one Merkle proof.
\begin{figure}[t]
    \centering
    \includegraphics[width=8cm]{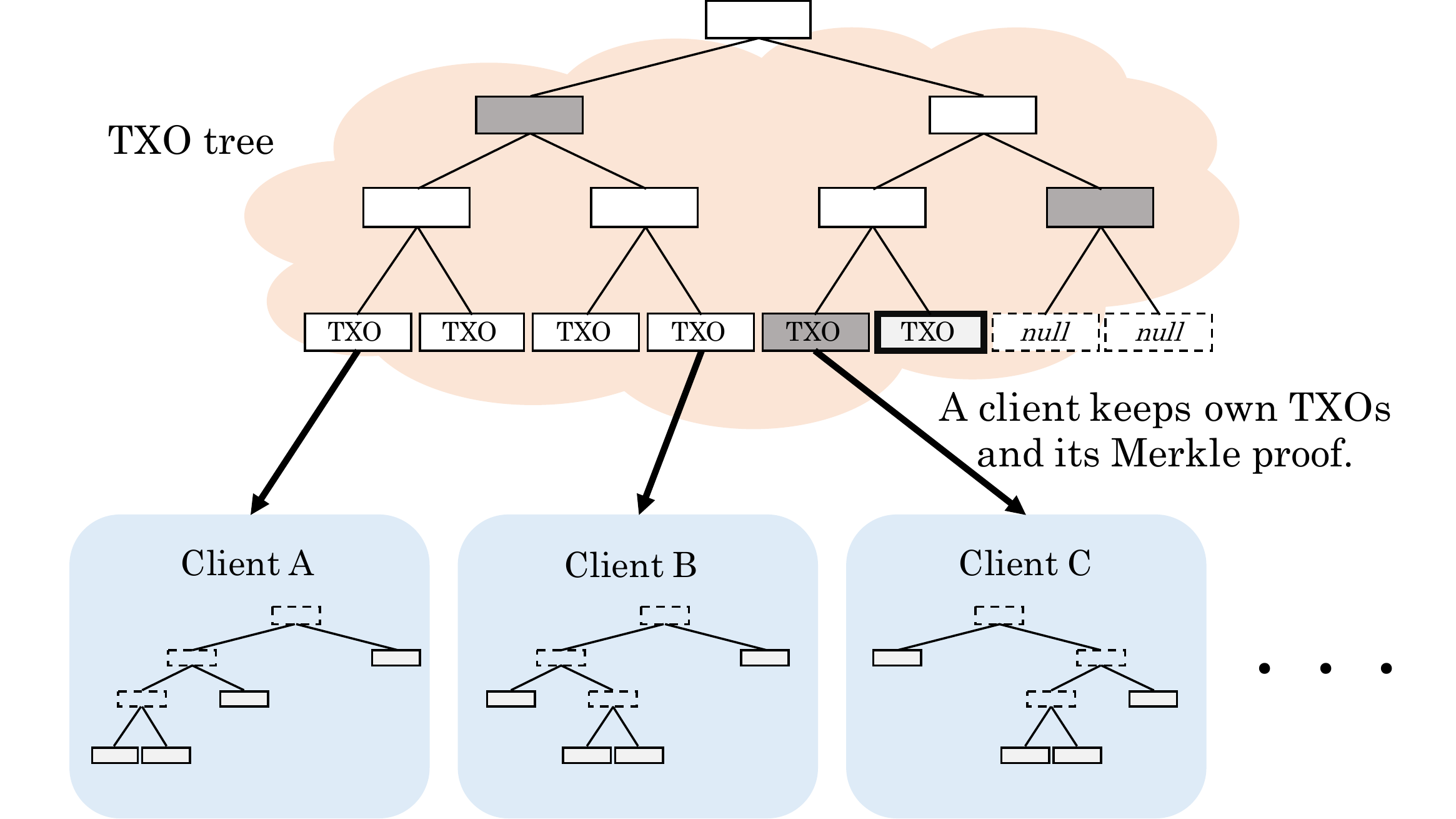}
    \caption{Concept of Trail architecture: The node with the thick border in the TXO tree is the rightmost leaf node, and grey nodes are Merkle proofs of the rightmost leaf node.}
    \label{fig:concept}
\end{figure}
As shown Fig.\ref{fig:concept}, the client holds only a part of the TXO tree related to its own TXO, and a Trail node holds only the root of the TXO tree, the hash value of the rightmost leaf node and its Merkle proof.

A TXO tree is a perfect binary Merkle tree, and the leaf nodes of TXO trees store the hash values of a TXO or a  null hash value.
When a new TXO is accepted, leaf nodes that are still null are assigned to TXOs from left to right.
A leaf node assigned a null value indicates that a TXO has not yet been assigned.
Since the new TXO is unused, the fixed-length hash value $\mi{hash}(\text{TXO})$ of the TXO is stored in the corresponding leaf node.
On the other hand, when a TXO is used, the corresponding leaf node contains the hash value $\mi{hash}(\text{TXO}^2)$, which is the concatenated value of the binary of the TXO.
$hash(b ^ 2)$ represents the hash value of the binary that concatenates two binaries of $b$.
If the binary of TXO is $n$ bytes, when TXO is used, the value obtained by multiplying the hash function by the binary of $2n$ bytes, that is, the concatenation of two TXO binaries, is stored in the leaf node.

The root of the TXO tree, the index of the rightmost leaf node to which a TXO is assigned, its hash value and Merkle proof are recorded in the block.

A node in the TXO tree is uniquely determined using the identifier called branchID.
The branchID is a concatenation of the height and index from the left in the height.
For example, if the height of the TXO tree is $2^8 = 256$, the branchID is 33 bytes, the first 1 byte represents the height of the node, and the remaining 32 bytes represent the index from the left in that height.
Let $\mi{branchID}(h, i)$ be the branchID of the $i$th node from the left in height $h$.

\subsection{Generation of a transaction by client}
\begin{table}[t]
    \caption{Data fields of transaction output (TXO).}
    \begin{center}
    \begin{tabular}{p{1.5cm}p{4.2cm}p{1.6cm}}
        \hline
        Field & Description & Size\\
        \hline
        Index & Index of corresponding leaf node. & 32 bytes\\
        ParentBlock & Hash value of the parent block of the block that added this TXO to the TXO tree. & 32 bytes \\
        OwnerAddress & The address of the owner of this TXO. & 32 bytes \\
        Balance & Amount of assets. & 32 bytes \\
        \hline
    \end{tabular}
\end{center}
    \label{tbl:txo}
\end{table}

\begin{table}[t]
    \caption{Data fields of transaction.}
    \begin{tabular}{p{1.5cm}p{4.2cm}p{1.6cm}}
        \hline
        Field & Description & Size \\
        \hline
        BlockHash & Hash value of the block that proofs of this transaction are based on. & 32 bytes\\
        Inputs & List of unused TXOs and their Merkle proofs. A Merkle proof is an array of 255 hash values. & \\
        Outputs & List of new TXOs. & \\
        Sigs & Signatures of accounts who own TXOs in Inputs or Outputs. & \\
        \hline
    \end{tabular}
    \label{tbl:tx}
\end{table}
The data structures of a TXO and transaction are shown in Table \ref{tbl:txo},\ref{tbl:tx}.

Clients keep their own TXOs and the update history of their Merkle proofs.
A transaction is generated by clients who consider the same block to be the latest block. 
Otherwise, Merkle proof validation will fail, and the transaction will not be contained in a new block.
A client can validate the Merkle proof of another client if it has the root of the TXO tree of the latest block.
For example, when checking whether a certain TXO is unused, a client checks whether the root value of the Merkle tree calculated from the hash value of the TXO and its Merkle proof matches the root value of the TXO tree of the latest block.
If it does not match, then the TXO has been used, the TXO has not yet been approved, or the Merkle proof is incorrect.

If all TXOs in Inputs are unused and the Merkle proof is correct, the client creates Outputs such that the total amount is less than or equal to the total amount of Inputs minus the fee.
At this time, no one knows which leaf node of the TXO tree the TXO will be assigned to, so the Index of TXOs in Outputs is empty.

Afterwards, the client broadcasts the transaction.

\subsection{Validating a transaction by Trail nodes}
Trail nodes validate received transactions and generate blocks from valid transactions.
Trail nodes validate the following 5 components:
\begin{itemize}
    \item Whether the block hash of the transaction matches the hash value of the latest block.
    \item Whether each TXO in the Inputs of the transaction is not in another transaction to include in new block.
    \item Whether the root of the TXO tree calculated from the Merkle proof in the Inputs and the hash value of TXO $\mi{hash}(\mi{TXO})$ is equal to the root of the latest block.
    \item Whether the total value of Outputs is less than or equal to the total value of Inputs minus fees.
    \item Whether the Index of the TXO in the Inputs is less than or equal to the RightmostIndex in the latest block.
\end{itemize}
The validator does not need all past blocks to validate if a TXO was previously used.

\subsection{Updating the Merkle proof in the transaction by Trail nodes}
A new block may be created and a TXO tree may be updated before a valid transaction is included in the block.
In this case, the Merkle proof in the transaction is no longer valid and cannot be included in the block.

For example, suppose two transactions containing the Merkle proof of the TXO tree of block $n$ are $T_1,T_2$.
If $T_1$ is contained in child block $n+1$ of block $n$, the node of the TXO tree will be updated.
Therefore, the root of the TXO tree of block $n+1$ and the root of the Merkle tree calculated from the TXO and its Merkle proof in transaction $T_2$ do not match because the root of the Merkle tree calculated from the TXO and its Merkle proof in transaction $T_2$ correspond to the root of the TXO tree of block $n$.
Therefore, $T_2$ cannot be included in block $n+2$ because it is not a valid transaction for block $n+1$.

Thus, for a pending transaction, the Trail node needs to update the Merkle proof in the transaction.
The proof can be updated with the information of the approved transaction in the blocks generated during the pending transaction
 because the hash value of the updated node of the TXO tree can be calculated from the information contained in those transactions.

The client does not include the Merkle proof when signing a transaction, as the Trail node may update the Merkle proof.

The update of the Merkle proof by Trail nodes is similar to the method proposed for Vault~\cite{b6}.

\subsection{Generating a block by Trail nodes}
\begin{table}[t]
    \caption{Data fields of a block.}
    \begin{tabular}{p{1.5cm}p{4.2cm}p{1.6cm}}
        \hline
        Field & Description & Size\\
        \hline
        Parent & Hash value of the parent block. & 32 bytes \\
        Root & Root of the TXO tree. & 32 bytes \\
        RightmostIndex & Index of the rightmost leaf node to which the TXO is assigned. & 32 bytes \\
        RightmostHash & Hash value of the leaf node corresponding to the RightmostIndex. & 32 bytes \\
        RightmostProof & Merkle proof of the leaf node corresponding to the RightmostIndex. & 255$\times$32 bytes \\
        \hline
    \end{tabular}
    \label{tbl:block}
\end{table}
The data structure of a block is shown in Table \ref{tbl:block}.
The block size is 8228 bytes, which is approximately  one hundredth of the block size in Bitcoin\cite{b13}.

In the block generating process, the Trail node computes the new root of the TXO tree.
First, hash values are assigned to leaf nodes (height 0) of the TXO tree in the following order: Merkle proof in Inputs, RightmostProof, RighmostHash, hash value of Outputs $\mi{hash}(\mi{TXO})$, and hash value of TXO in Inputs $\mi{hash}(\mi{TXO}^2)$.

The TXOs in Outputs of all valid transactions are assigned in order from the leaf node of the RightmostIndex+1 without gaps.
At this time, the Index of the TXO in Outputs is fixed. 
Therefore, the hash value of the TXO in Outputs including the Index is assigned to a leaf node.
Since the TXOs in Inputs are used, the hash value flagged as used $\mi{hash}(\mi{TXO}^2)$ is assigned to the corresponding leaf node.
Furthermore, RightmostHash is assigned to the leaf node whose index is RightmostIndex.

Note that hash values are assigned to only a portion of the leaf nodes.
The block proposer computes the hash value of the parent node from the hash values assigned to leaf nodes.
The hash value is always assigned to the sibling nodes of the leaf nodes to which the hash value is assigned, except the rightmost leaf node.
If the hash value is not assigned to the sibling node of the rightmost node, the hash value of the parent node is computed assuming that a null hash value is assigned to the sibling node.

Then, hash values are assigned to the node at height 1 in the following order: Merkle proof in Inputs, RightmostProof, and the hash value computed from the leaf nodes.
Sibling nodes of nodes to which hash values have been assigned, except the rightmost node, are always assigned hash values.
If a hash value is not assigned to the sibling node of the rightmost node, $\mi{hash}(\mi{hash}(\mi{null})^2)$ is assigned, and the Trail node computes the hash value of the parent node.

Similarly, the hash value is assigned to the node at height $h+1$ in the following order: Merkle proof in Inputs, RightmostProof, and the hash value computed from the nodes at height $h$.
Additionally, the block proposer computes the hash value of the node in height $h+1$.
Finally, the block proposer obtains a new root of the TXO tree.

The block proposer generates a new block with the index of the rightmost leaf node to which the newly added TXO is assigned as the RightmostIndex, the hash value of that node as RightmostHash, and the Merkle proof of that node as RightmostProof.
Then, the block proposer broadcasts the new block and the approved transaction to other Trail nodes.

\begin{figure*}[t]
    \centering
    \includegraphics[width=15cm]{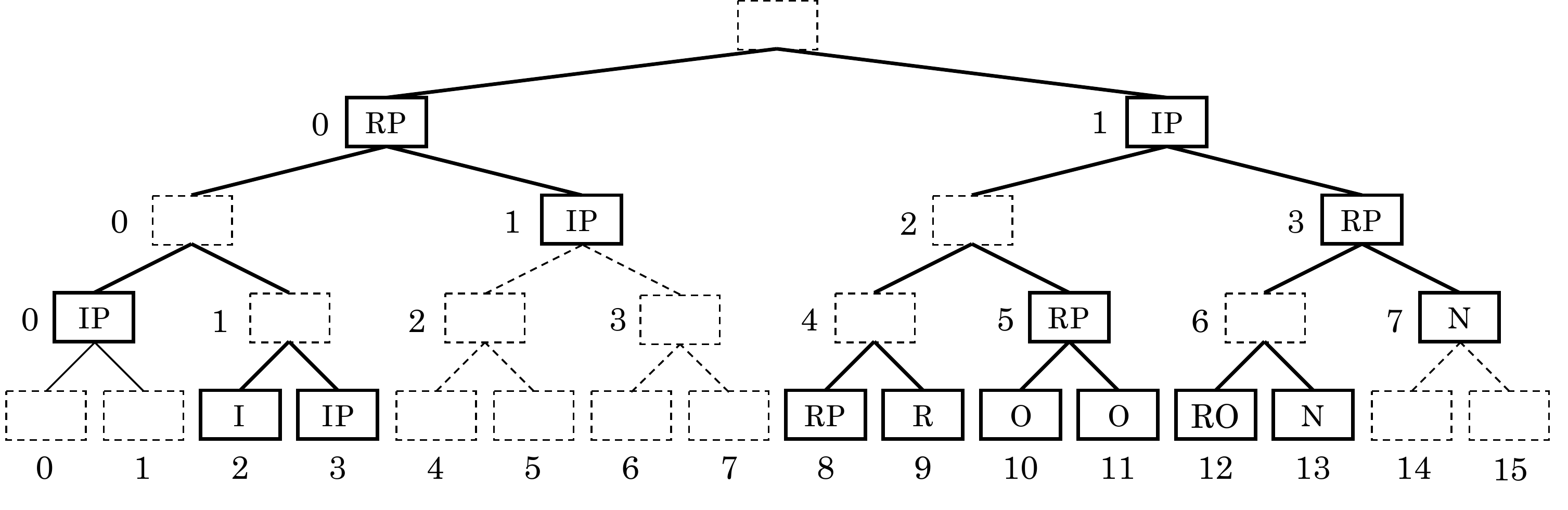}
    \caption{Example of assigning a hash value to a TXO tree from transaction and parent block data:
    Node I is assigned the hash value of TXO in Inputs. 
    Node IP is assigned the hash value of the Merkle proof in Inputs. 
    Node R is assigned the RightmostHash of the parent block. 
    Node RP is assigned the RightmostProof of the parent block. 
    Node O is assigned the hash value of the TXO in Outputs. 
    Node RO is the rightmost node, which is assigned the hash value of the TXO in Outputs. 
    Node N is assigned a null hash value.
    Node in the dashed box is assigned no hash value initially.
    The block proposer computes the hash value of the parent node along the thick line between nodes.
    The numbers beside nodes are the indexes of the nodes in that height.}
    \label{fig:txo-tree}
\end{figure*}

In Fig.\ref{fig:txo-tree}, the leaf nodes at indexes 2, 3, 8, 9, 10, 11, 12, and 13 are assigned hash values.
The node at index 0 is assigned the hash value of the TXO in Inputs $\mi{hash}(\mi{TXO}^2)$, and the node at index 3 is assigned the Merkle proof in Inputs.
The node at index 8 is assigned the RightmostProof, and the node at index 9 is assigned the RightmostHash;
that is, the RightmostIndex of parent block is 9.
The nodes at indexes 10, 11, and 12 are assigned the hash value of Outputs $\mi{hash}(\text{TXO})$.
Since the node at index 12 is the rightmost node assigned a hash value, the RightmostIndex of the new block will be 12.

The sibling nodes of the nodes at indexes 2, 3, 8, 9, 10, and 11 are assigned hash values;
however, the sibling nodes of the node at index 12 have not yet been assigned a TXO.
Therefore, the node at index 13 is assigned a null hash value.

At this time, since the sibling nodes of the leaf node to which hash values have been assigned have all been assigned a hash value, the block proposer can compute the hash values of the parent nodes.

Then, the block proposer assigns hash values to nodes at height 1.
The node at index 0 is assigned the Merkle proof in Inputs, and the node at index 5 is assigned the RightmostProof.
Furthermore, the nodes at indexes 1, 4, and 6 are assigned hash values computed from the leaf nodes.
The sibling nodes of the nodes at indexes 0, 1, 4, and 5 are also assigned a hash value.
However, since the sibling nodes of the nodes at index 6 have not been assigned a hash value, the node at index 7 is assigned a null hash value $\mi{hash}(\mi{hash}(\mi{null})^2)$.

At this time, since the sibling nodes of the node at height 1 to which a hash value has been assigned have all been assigned a hash value, the block proposer can compute the hash values of the parent nodes.

The nodes at heights 2 and 3 are assigned hash values in the same way.
The hash values of nodes other than the one to which the hash value is assigned are not updated when computing the new root of the TXO tree.

In this case, the new RightmostIndex is 12, the new RightmostHash is the hash value of the node at $\mi{branchID}(0,12)$, and the new RightmostProof is the hash values of the nodes at $\mi{branchID}(0,13)$, $\mi{branchID}(1,7)$, $\mi{branchID}(2,2)$, and $\mi{branchID}(3,0)$.

\subsection{Updating the client's data}
Finally, we describe the procedure for updating the data of the client. 
When a block is created, the client receives the updated hash value of the node in the TXO tree to update the Merkle proof it has.

Here, it is assumed that there is no fork or churn, and the message does not disappear during communication with peers.
That is, the client receives all updates for the blocks and the TXO tree:
we consider the case where the client cannot receive a portion of the updates in Section \ref{sec:sync}.

The client receives the new block, the new TXOs, the used TXOs, and the hash values of the nodes that have been assigned hash values when computing the new root of the TXO tree.
If the client has its own TXO in the new TXOs, the client keeps that TXO as an unused TXO.
If the client has its own TXO in the used TXOs, the client marks the TXO as used.

If the Merkle proof of the unused TXOs is updated or the client has not yet obtained it, the client recodes the hash value of the Merkle proof together with the branchID and the hash value of the new block.

\section{Data synchronization}
\label{sec:sync}
In the previous section, we assumed there were no forks and that the client was always connected to the network.
However, in an actual blockchain network, forks occur, and the client repeatedly connects to and disconnects from the network.
In this case, the client needs to obtain the blockchain data and synchronize its own data to the blockchain.

Further, a Trail node may have newly joined or left the network and have not obtained some blocks.
In this case, the node needs to obtain the blockchain data and synchronize its own data.

\subsection{Full node}
For data synchronization, a full node with all the transactions is required.
The full node is a subclass of Trail nodes. Normal Trail nodes do not need to hold transactions approved in past blocks, and
Trail nodes do not have to be a full node for block generation and transaction validation.
A full node requires more storage than normal Trail nodes but approximately the same storage as a full node in Bitcoin.

When a fork or churn occurs, the client obtains TXO tree updates that have not been obtained from the full node.
Algorithm \ref{algo:download-txo-tree-update} shows how the client obtains TXO tree updates from a full node.
It is sufficient for the client to obtain only the latest update for each node of the TXO tree related to its own TXOs rather than all the updates.
\begin{algorithm}[t]
    \small
    \caption{Download TXO tree latest updates from full node.}
    \label{algo:download-txo-tree-update}
    \SetAlgoLined
    \KwIn{List of branchID $\mi{ids}$, Hash value of block $\mi{from}$, $\mi{to}$}
    \KwOut{Hash map the hash value of nodes to branchID and hash value of the block}
    \SetKwBlock{Beginn}{beginn}{ende}
    \Begin{
        \uIf{$\mi{from}$ is not ancestor of $\mi{to}$}{
            \KwRet{error}       
        }
        \Else{
            $\mi{nodes} \leftarrow$ nodes which branchID is in $\mi{ids}$\;
            \KwRet latest updates of $\mi{nodes}$ between $\mi{from}$ and $\mi{to}$
        }
    }
\end{algorithm}

The Trail node requests the transactions approved by the missing block from the full node.
The block itself can be obtained from other Trail nodes that are not full nodes, but the Trail node needs to obtain the transaction from the full node to validate missing blocks.

\subsection{Data synchronization for client}

Table \ref{tbl:client-data} shows the fields of the data held by the client.
The client keeps the hash value of blocks received, own used TXOs, own unused TXOs, and update history of Merkle proofs of TXOs.
\begin{table}[t]
    \caption{Data fields of client.}
    \begin{tabular}{p{2cm}p{6cm}}
        \hline
        Field & Description \\
        \hline
        LatestBlock & Hash value of the latest block. \\
        Blocks & Hash map that maps the hash value of the received blocks to the hash value of the parent block. \\
        Unused & Hash map that maps the hash value of the latest block of each fork to the list of unusedTXO at that block. \\
        Used & Hash map that maps the TXOs to the hash values of blocks with used TXOs. \\
        Updates & Hash map that maps the hash value of nodes to branchID and hash value of the block. \\
        \hline
    \end{tabular}
    \label{tbl:client-data}
\end{table}
When a new block is generated, client $c$ receives new block $b$, used TXO $\mi{usedTXOs}$, newly added TXO $\mi{newTXOs}$, and node hash values $\mi{nodeHashes}$.
When a client receives a new block, the client first downloads missing blocks and TXO tree updates from the full node.
Then, the client updates its own data with $b$, $\mi{usedTXOs}$, $\mi{newTXOs}$, and $\mi{nodeHashes}$.

\section{Broadcast}
Since transactions in Trail include the TXO Merkle proof, the data size of the transaction is large and wastes network resources.
This section describes a technique for reducing the data size of a transaction using the characteristics of the TXO tree.
Furthermore, during block propagation, the data size to be broadcast is reduced using the existing blockchain method.
Finally, we describe the process through which a client obtains the new block, TXOs, and TXO tree updates.

\subsection{Transaction}
A transaction includes the TXOs and the Merkle proof in Inputs.
For each TXO, the size of the transaction increases by $128 + 32 \times 255 = 8288$ bytes.
Therefore, Trail attempts to reduce the number of TXOs in Inputs by determining the transaction fee according to the number of TXOs in Inputs.
To reduce the Inputs, the client manages the balance with only a small number of TXOs instead of keeping the balance in a large number of small TXOs.
Assuming the client generates only one transaction per one block, the client needs to have only one unused TXO.

Furthermore, multiple Merkle proofs included in a transaction can be combined to reduce the data size of the transaction.
Assume a client generates a transaction at least once at $t$, and the transaction is included in a block.
Let $\mi{interval}$ be the block generation interval: the number of blocks generated per $t$ is $\frac{t}{\mi{interval}}$.
Further, assuming that the number of TXOs added per block is $n$, the difference between the maximum index and the minimum index of TXOs in Inputs is $\frac{tn}{\mi{interval}}$. 
Therefore, all nodes that are higher than the height $\log_2 \frac{tn}{\mi{interval}}$ of the Merkle proof in Inputs are the same.

For example, in the case of $t = 7\text{ day} = 604800\text{ sec}$, $interval = 15\text{ block/sec}$, $n = 10000\text{ tx/block}$, $\log_2 \frac{tn}{\mi{interval}}$ is less than 29.
Therefore, the increase in the size of the transaction per TXO in Inputs is $128 + 29\times32 = 1056$ bytes.

Furthermore, the node at height higher than $\log_2 \frac{tn}{\mi{interval}}$ of the Merkle proof in Inputs is the same as the node with a height higher than $\log_2 \frac{tn}{\mi{interval}}$ of the RightmostProof of the latest block.
Therefore, the Merkle proofs at height $\log_2 \frac{tn}{interval}$ or more can be removed from the transaction by adding a flag to the transaction.

The size of the transaction is then $32+i\times 1056 + o\times128$ bytes, where the number of TXOs in Inputs is $i$ and the number of TXOs in Outputs is $o$.

\subsection{Block}
For block validation, the block proposer must broadcast the approved transactions with the new block;
thus, the size of the data to be broadcast is substantial.
If the total number of TXOs in Inputs approved by the block is 10000 and the total number of Outputs is 10000, the data size will be approximately $(960+128)\times10000$ bytes $\approx$ 10 MB.

The data size can be reduced by omitting duplicate Merkle proofs, but it is expected that this problem can be solved  using a protocol similar to compact block relay\cite{b9}.
Compact block relay is a Bitcoin protocol that includes only transaction IDs in the block broadcast data instead of sending entire transactions because the transactions are broadcast before the block is broadcast and other nodes in the network already have the transactions.
By means of compact block relay, the transactions included in the block broadcast can be compressed to 8 bytes, so even if 10,000 transactions are approved, the data size is, at most, block size $+ 10000 \times8$ = 8288 + 80000 bytes $\approx$ 90 KB.

\subsection{New TXOs, used TXOs, TXO tree updates}
When a block is generated, the client needs to receive the newly added TXOs, used TXOs, and TXO tree updates to update its own data.
However, the client needs to update only its own TXOs and Merkle proof;
thus, the client does not need all these data.

In Trail, when the client receives a new block, the client sends a message to the node containing the hash value of the block, the address of the client, and the indexes of its own TXOs.
The Trail node returns the newly added TXOs or used TXOs whose OwnerAddress matches the address in the message and the hash values of nodes with the branchID in the message.

\section{Data Archiving}
Trail assumes clients use mobile devices. 
Therefore, the client does not keep unnecessary data on the device but archives the data to external storage, such as the cloud or SSD.

Here, only the Merkle proofs of unused TXOs at the latest block are stored on the device, and the hash values of other nodes are archived.
Instead of $c.\mi{Updates}$, the update data on the device of client $c$ is represented by $c.\mi{Memory}$, and the archived update data is represented by $c.\mi{Archive}$.
Algorithm \ref{algo:update-client-data-with-archive} illustrates the archiving of unnecessary data to $c.\mi{Archive}$.
\begin{algorithm}[h]
    \small
    \caption{Update the update history with archiving.}
    \label{algo:update-client-data-with-archive}
    \SetAlgoLined
    \DontPrintSemicolon
    \KwIn{client $c$, block $b$}
    \Begin{
        $\mi{newMemory} \leftarrow$ HashMap\;
        \For{$t$ in $c.\mi{Unused}[\mi{hash}(b)]$}{
            $\mi{index} \leftarrow t.\mi{Index}$\;
            \For{$h \leftarrow 0$ to $254$}{
                \uIf{$\mi{index}$ is even}{
                    $\mi{index} \leftarrow \mi{index}+1$
                }
                \Else{
                    $\mi{index} \leftarrow \mi{index}-1$
                }
                $i \leftarrow \mi{branchID}(h,\mi{index})$\;
                \uIf{$c.\mi{Memory}[i]$ exists}{
                    $\mi{updates} \leftarrow c.\mi{Memory}[i]$\;
                    \If{$\mi{nodeHashes}[i]$ exists}{
                        $\mi{newHash} \leftarrow \mi{nodeHashes}[i]$\;
                        $\mi{updates}[hash(b)] \leftarrow \mi{newHash}$
                    }
                }
                \uElseIf{$\mi{nodeHashes}[i]$ exists}{
                    $\mi{newHash} \leftarrow \mi{nodeHashes}[i]$\;
                    \uIf{$c.\mi{Archive}[i]$ exists}{
                        $\mi{updates} \leftarrow c.\mi{Archive}[i]$\;
                        $\mi{updates}[\mi{hash}(b)] \leftarrow \mi{newHash}$
                    }
                    \Else{
                        $\mi{updates} \leftarrow$ HashMap\;
                        $\mi{updates}[\mi{hash}(b)] \leftarrow \mi{newHash}$
                    }
                }
                \Else{
                    $\mi{updates} \leftarrow c.\mi{Archive}[i]$
                }
                $\mi{newMemory}[i] \leftarrow \mi{updates}$\;
                $\mi{index} \leftarrow \mi{index} \gg 1$
            }
        }
        \ForAll{$i \in c.\mi{Memory.keys}$}{
            \If{$\mi{newMemory}[i]$ not exists}{
                Store $c.\mi{Memory}[i]$ to $c.\mi{Archive}[i]$
            }
        }
        $c.\mi{Memory} \leftarrow \mi{newMemory}$
    }
\end{algorithm}

Further, the client archives old updates of the TXO tree.
Let $h_{\mi{latest}}$ be the block height of the latest block and $h_{\mi{archive}}$ be the threshold for archiving.
The client archives the updates of the TXO tree in blocks with a block height of less than $h_{\mi{latest}} - h_{\mi{archive}}$.

Let $u$ be the number of unused TXOs for the latest block of each fork and $f$ be the number of forks when generating $h_{\mi{archive}}$ blocks.
Assume that unused TXOs are used during $b$ blocks, that $n$ TXOs are added for each block, and that the Merkle proofs of all unused TXOs are updated in every block.

At this time, in each fork, the Merkle proofs of the unused TXOs with height larger than $\log_2 bn$ are the same.
Therefore, the data size of the TXO tree updates that the client keeps on the device is $\min(h_{\mi{archive}},b) \times 32f ( u\log_2 bn + 255-\log_2 bn)$ bytes.
Additionally, the data size of unused TXOs is $128uf$ bytes.
In the case of $(u = 1, h_{\mi{archive}} = 100, b = 40320, f = 2, n=10^4)$, the data size on the device is $\min(h_{\mi{archive}},b) \times 32f ( u\log_2 bn + 255-\log_2 bn) + 128uf \approx$ 1.63 MB.
By properly archiving the data on the device, the total size can be reduced to approximately 1.63 MB, regardless of the length of the blockchain.
This amount of data is sufficiently small to store on a mobile device.

On the other hand, the size of archived data increases as the blockchain lengthens.
However, the data size can be reduced by deleting archived data at block height $h_{\mi{delete}}$. 
The client considers that blocks whose height is lower than $h_{\mi{latest}}-h_{\mi{delete}}$ are finalized and will not be overwritten and deletes the update history in those blocks.
If a block is overturned by an attack, the client has to obtain the deleted data from the full node.
$h_{\mi{delete}}$ is considered to be larger than $h_{\mi{archive}},b$.
The size of archived TXO tree updates is at most $32\frac{h_{\mi{delete}}f}{b}(ub\log_2 bn + \frac{h_{\mi{delete}}}{b}(255-\log_2 bn))$ bytes.
Moreover, the size of archived TXOs is $128(\frac{h_{\mi{delete}}}{b})^2uf$ bytes.
In the case of $({u = 1}, {h_{\mi{archive}} = 100}, {b = 40320}, {f = 2}, {n=10^4},h_{\mi{delete}} = 10^5)$, the size of archived data is approximately 183 MB.

\section{Conclusion}
We proposed Trail architecture to reduce the storage size of nodes.
Trail manages the assets of the account by updating one Merkle tree through the blockchain.
The transaction issuer includes its own TXO and its Merkle Proof in the transaction, which
allows the node to validate and generate blocks without having to hold the entire Merkle tree.

We described techniques to reduce the data size broadcast to the network.
The transaction data size can be reduced by omitting duplicate Merkle proofs.
Furthermore, during block propagation, the data size is reduced using compact block relay.

Finally, we show that by properly archiving and deleting data, the data on the client device can be reduced to approximately 1.6 MB, and the data to be archived can be reduced to 183 MB.
Therefore, Trail works on mobile devices.

\section*{Acknowledgment}
This work was supported by SECOM Science and Technology Foundation.

\end{document}